\documentstyle[preprint,aps]{revtex}
\begin{document}
\draft
\title{Energy Gap Induced by Impurity Scattering:\\
New Phase Transition in Anisotropic Superconductors}
\author{ S. V. Pokrovsky$^1$ and V. L. Pokrovsky$^{2,3}$}
\address{$^1$Department of Physics, Massachusetts Institute of
Technology\\
Cambridge, MA, 02139-4307\\
$^2$Department of Physics, Texas A\&M University,\\
College Station, TX 77843-4242*\\
$^3$Landau Institute for Theoretical Physics,\\
Kosygin St. 2, Russia, Moscow 117940}
\date{\today}
\maketitle
\begin{abstract}
It is shown that layered superconductors are subjected to a phase
transition at zero temperature provided the order parameter (OP)
reverses its sign on the Fermi-surface but its angular average
is finite.
The transition is regulated by an elastic impurity scattering rate $1/\tau$.
The excitation energy spectrum, being gapless at the low level of scattering,
develops a gap as soon as the scattering  rate exceeds some critical value
of $1/\tau_\star$.
\end{abstract}
\pacs{PACS numbers:  74.20.Fg, 74.25.Jb, 74.62.-c, 74.72.Bk}
\narrowtext
It has been shown recently that non-magnetic impurities may destroy the
$d$-wave superconductivity in the same manner as magnetic
impurities do it in the $s$-wave conventional superconductors
\cite{radtke,pines,borkovski}.
However, this pure $d$-wave scenario is not supported by the existent
experimental data, at least in $YBaCuO$ samples, subjected to radiation
damage or doping \cite{residual}. One might anticipate that $T_c$
should drop abruptly above a certain critical concentration of the defects.
Instead,  critical temperature has been found just to decrease gradually in
the whole range of a residual resistivity variation.
The above discrepancy has been considered as a compelling counter-argument
against the whole idea of exotic pairing in this compound.
However, in this note we make the case
that the above mentioned experimental results do
not in fact contradict the alleged non-trivial structure of the
order parameter (OP),
provided the latter is determined by the symmetry of the crystal
and, hence, deviates from the exact $d$-wave form. In this situation,
as will be demonstrated below, an initially gapless excitation
spectrum of a clean superconductor may acquire a gap due to a
scattering of electrons by non-magnetic impurities.  An elastic
scattering gives rise to this paradoxical effect, once the OP in the
clean material possesses the nodes on the Fermi surface whereas its
angular average still does not vanish, and the scattering rate exceeds
some critical value.
\par
The results of recent Josephson tunneling
experiments \cite{Josephson} and the measurement of critical
temperature versus residual resistivity variation \cite{residual} may
be interpreted as an evidence in favor of this very structure of the OP.
Indeed, the Josephson tunneling data in the corner geometry imply the
sign reversal of the OP on the Fermi surface. On the other hand, as
is already mentioned, the growth of the residual resistivity is
accompanied by a slow decrease of critical temperature.  This type of
behavior rules out the nullification of the OP averaged over the Fermi
surface as we shall show later.
\par
It should be noted that recent measurements of a Josephson current in
$c$-direction by Sun {\it et al.} \cite{sun} do not conform easily
with the phase alternating OP. The values of the net current found
in this experiment for the heavily twinned samples in plain geometry are
several orders of magnitude higher than theoretically estimated.
The interpretation of these results may prove to be a subtle matter,
however. A possible source of the finite
Josephson current could be a symmetry violation caused by surface
defects and the surface itself together with the phase
self-adjustment. For other plausible scenarios this effect
see preprints \cite{Kuboki}.
For now we assume that the OP sign reversal, as well as the non-zero
value of $\langle \Delta \rangle$, is reliably established in the
experiments cited above \cite{Josephson,residual}.
\par
We analyze the problem in the framework of the anisotropic BCS model.
The OP is determined by self-consistency equation:
\begin{equation}
\Delta(\phi ) \,=\, T\sum_n\int V(\phi,\phi^{\prime})
\frac{\tilde{\Delta}_n(\phi^{\prime})}{\sqrt{\tilde{\eta}_n^2 +
\tilde{\Delta}_n^2(\phi^{\prime})}}\frac{d\phi^{\prime}}{2\pi}
\label{eq:self-consistency}
\end{equation}
Here the summation goes over the Matsubara imaginary frequencies
whereas the integration is restricted to the Fermi surface, which is
presumed to be a cylinder. The kernel $V(\phi,\phi^{\prime})$
describes the interaction of electrons on the Fermi surface.
The renormalized Matsubara frequency $\tilde{\eta}_n$ and the
frequency-dependent OP $\tilde{\Delta}_n(\phi)$ are related to the
corresponding bare quantities $\eta_n=(2\,n+1)\,\pi\,T$ and
$\Delta(\phi)$ via Abrikosov-Gor'kov (AG) equations:
\begin{eqnarray}
\tilde{\eta}_n\, -
\,\frac{\tilde{\eta}_n}{\tau}
\left \langle\frac{1}
{\sqrt{\tilde{\eta}_n^2+\tilde{\Delta}_n(\phi )^2}}
\right \rangle
&=&\, \eta_n
\label{eq:stepsepsilon}\\
\tilde{\Delta}_n(\phi )\,- \,\frac{1}{\tau}
\left \langle
\frac{\tilde{\Delta}_n(\phi^{\prime} )}
{\sqrt{\tilde{\eta}_n^2+\tilde{\Delta}_n(\phi^{\prime} )^2}}
\right \rangle
&=&\,\Delta(\phi )
\label{eq:stepsdelta}
\end{eqnarray}
The angular brackets in the AG equations denote the angular average.
For the sake of simplicity we consider an isotropic scattering only.
For the same reason we have disregarded any dependence of the OP and
the electronic interaction potential on the $c$-component of momentum.
\par
Let us first examine the behavior of the critical temperature with
respect to variation of the scattering rate. In the vicinity of
critical line the equations
(\ref{eq:self-consistency}-\ref{eq:stepsdelta})
may be linearized and the infinitesimal order parameter may be
eliminated.  The result reads:
\begin{equation}
	1\,=\,g(T,\tau)\,\left\langle\left(1\,-
		\,f(T,\tau)\hat{V}\right) ^{-1} \bar{V}(\phi)
			\right\rangle
\label{eq:T-c-formal}
\end{equation}
Here we defined a linear operator $\hat{V}$ related to the electronic
interaction kernel:
\begin{equation}
	\hat{V}\, \psi(\phi)\, =
		\, \int V(\phi ,\phi^{\prime})\, \psi(\phi^{\prime})
			\frac{d\phi^{\prime}}{2\pi}
\label{eq:kernel-def}
\end{equation}
This operator maps $1$ onto the function $\bar{V}(\phi)$
\begin{equation}
	\bar{V}(\phi)\,=
		\,\int V(\phi ,\phi^{\prime})\frac{d\phi^{\prime}}{2\pi}
\label{eq:1-mapping}
\end{equation}
The functions $f(T,\tau )$ and $g(T,\tau )$ in eqn. (\ref{eq:T-c-formal})
are defined by the
following expressions:
\begin{eqnarray}
f(T,\tau )\,=\,T\sum_n\frac{1}{\mid\eta_n\mid + 1/\tau}
\label{eq:fdef}\\
g(T,\tau )\,=\,T\sum_n\frac{1}{\mid\eta_n\mid(1+\mid\eta_n\mid\tau)}
\label{eq:gdef}
\end{eqnarray}
The summation in the expression for $f(T,\tau )$ is limited by an
ultraviolet cutoff: $\eta_n < \bar{\epsilon}$. The latter is related to
the critical temperature $T_{c0}$ of the clean superconductor as follows:
\begin{equation}
	T_{c0} = \frac{2 \gamma \bar{\epsilon}}{\pi} \exp(-\pi/V_0)
\label{T-c-clean}
\end{equation}
where $V_0$ is the maximal eigenvalue of the operator $\hat{V}$ and
$\gamma$ is the Euler constant. In the dirty limit $(\tau T_{c0} \ll 1)$
the approximate solution of eqn. (\ref{eq:T-c-formal}) reads
\begin{equation}
T_c\,=\,\gamma\,\bar{\epsilon}\,
	(\gamma\,\bar{\epsilon}\,\tau/2)^{\kappa- 1}
		  \exp (-\pi /\langle\bar{V}(\phi)\rangle )
\label{eq:T-c-dirty}
\end{equation}
where
$
	\kappa = \langle \bar{V}^2 \rangle/\langle \bar{V} \rangle^2
$.
Within the same approximation
$
	\Delta(\phi) \propto \bar{V}(\phi)
$.
Hence, the exponent $\kappa$ may be expressed in
terms of the OP:
$
	\kappa = \langle \Delta^2 \rangle / \langle\Delta \rangle^2
$.
The power law (\ref{eq:T-c-dirty}) has been derived in the sixties by
P.~Hohenberg \cite{hohenberg} under the assumption of a weak anisotropy.
We have found that it is the high scattering level that really matters.
Notice that the exponent $\kappa$ becomes infinite when
$
	\langle \Delta \rangle \rightarrow 0
$.
A more scrupulous analysis \cite{radtke,pines,borkovski} shows that if
$\langle \Delta \rangle = 0$
the critical temperature vanishes at some finite scattering rate
proportional to the critical temperature of the clean superconductor:
$
	2\,\pi\,\tau_c\,T_{c0} =1
$.
\par
Equation (\ref{eq:T-c-dirty}) might be interpreted directly as the relation
between the critical temperature and the residual resistivity:
$
	T_c(\rho) \propto \rho^{\kappa-1}
$,
provided that the effective number of carriers does not depend
substantially on the concentration of defects.
\par
Let us study equations
(\ref{eq:self-consistency}-\ref{eq:stepsdelta})
at temperature equal to zero. In this case the summation in
eqn. (\ref{eq:self-consistency}) should be replaced by
integration over the continuous variable $\eta$ and the domain of
functions $\tilde{\eta}(\eta)$ and $\tilde{\Delta}(\phi,\eta)$ extends
into the whole positive half-axis of the same variable.
We are interested in the behavior of the above functions at $\eta
\rightarrow 0$. The values $\tilde{\eta}(0)$ and
$\tilde{\Delta}(\phi,0)$ determine the density of states (DOS) on the Fermi
surface, vanishing if $\tilde{\eta}(0) = 0$.
\par
We are going to prove the following three statements
concerning the DOS on the Fermi surface:
\begin{enumerate}
	\item Let $\langle \Delta(\phi) \rangle \neq 0$ and
		$\Delta(\phi)$
		has no nodes. Then
		$\tilde{\eta}(0) = 0$ for any $\tau$.
	\item Let $\langle \Delta(\phi) \rangle = 0$.
		Then
		$\tilde{\eta}(0) > 0$
		for any
		$\tau \geq (2\pi T_{c0})^{-1}$.
	\item Let $\langle \Delta(\phi) \rangle \neq 0$
		but $\Delta(\phi)$ possesses nodes. Then
		$\tilde{\eta}(0) = 0$ for $\tau < \tau_\star$
		but
		$\tilde{\eta}(0) > 0$ for $\tau > \tau_\star$.
		The equation for $\tau_\star$ as a functional of
		$\Delta(\phi)$
		will be derived below.
\end{enumerate}
Before proceeding to the proof let us remark that according
to eqn. (\ref{eq:stepsepsilon}) the following separation of variables
takes place:
\begin{equation}
	\tilde{\Delta}(\phi,\eta) = \Delta(\phi) + \sigma(\eta)
\label{eq:sigma-def}
\end{equation}
and the {AG} equations may be rewritten in terms of the function
$\sigma(\eta)$ just defined:
\widetext
\begin{eqnarray}
	\tilde{\eta}(\eta) \left( 1-\frac{1}{\tau}\left \langle
		\frac{1}{\sqrt{(\Delta(\phi) +\sigma(\eta))^2 +
			\tilde{\eta}(\eta)^2}}
	\right \rangle\right) &=& \eta
\label{eq:AG-eta}\\
	\sigma(\eta)\left( 1-\frac{1}{\tau}\left \langle
	\frac{1}{\sqrt{(\Delta(\phi) +\sigma(\eta))^2 +
		\tilde{\eta}(\eta)^2}}
	\right \rangle\right) &=&
	\frac{1}{\tau}\left \langle\frac{\Delta(\phi)}
	{\sqrt{(\Delta(\phi) +\sigma(\eta))^2 +\tilde{\eta}(\eta)^2}}
	\right \rangle
\label{eq:AG-sigma}
\end{eqnarray}
\narrowtext
The first proposition stems from eqns. (\ref{eq:AG-eta},\ref{eq:AG-sigma})
straightforwardly.  Really, let $\tilde{\eta}(0) \neq 0$, then the
expression in the brackets vanishes and, consequently, the r.h.s of
eqn. (\ref{eq:AG-sigma}) must have a root at $\eta = 0$. The latter is
impossible, however, if $ \Delta(\phi)$ does not reverse its sign.
Hence, if $\Delta(\phi)$ has no nodes $\tilde{\eta}(0) = 0$.
\par
Let us skip the second statement for a moment and consider the case
when the OP in a clean superconductor does reverse its sign but its
angular average is finite.
\par
We study, first, the limit of small impurity concentration $(\tau
\left \langle \Delta(\phi) \right \rangle \gg 1)$.  Collecting
terms proportional to $1/\tau$ in eqn. (\ref{eq:AG-sigma}) one can find out
that $\sigma = O(1/\tau)$. In other words, renormalization of the order
parameter is small. In particular, its nodes remain almost at their
original locations. However, renormalization of $\tilde{\eta}(\eta)$ is
not small due to the logarithmic divergence of the term,
proportional to $1/\tau$ in the l.h.s. of eqn. (\ref{eq:AG-eta}) at
$\tilde{\eta}(\eta) = 0$.  A closer examination of the l.h.s of eqn.
(\ref{eq:AG-eta}), considered as the function $\eta(\tilde{\eta})$
(inverse to $\tilde{\eta}(\eta)$), shows that it departs from the
coordinate origin with an infinite negative derivative and,
after reaching its minimum, crosses the abscissa axis at
$\tilde{\eta}_0 = \Delta_L^\prime \exp{(-\tau \Delta_I^\prime /2)} $,
where
$1/\Delta_I^\prime=(1/\mid\Delta_1^\prime\mid +
				1/\mid\Delta_2^\prime\mid)$,
whereas
$               \log{\Delta_L^\prime}=
(\mid\Delta_1^\prime\mid        \log{\mid\Delta_1^\prime\mid}+
\mid\Delta_2^\prime\mid\log{\mid\Delta_2^\prime\mid})/
(\mid\Delta_1^\prime\mid+\mid\Delta_2^\prime\mid) $
and $\Delta_{1,2}^\prime$ denote the derivatives of the order
parameter at its nodes.  Among the two roots, only $\tilde{\eta}_0$
resides on the physical sheet\cite{remark}.
The first part of the statement 3 is proved.
\par
In the opposite limit of large scattering rate $(1/\tau \gg T_{c0})$ the
asymptotic solution of eqns. (\ref{eq:AG-eta},\ref{eq:AG-sigma}) can be
found in the following form:
\begin{eqnarray}
\tilde{\eta}(\eta) &=&  \eta \left( 1 + \frac{1}{\tau
\sqrt{\left \langle \Delta(\phi) \right \rangle^2 +\eta^2}}\right)
\label{asympt-eta} \\
\sigma(\eta) &=&
\frac{\left \langle \Delta(\phi) \right \rangle}{\tau
\sqrt{\left \langle \Delta(\phi) \right \rangle^2 +\eta^2}}
\label{asympt-sigma}
\end{eqnarray}
Plugging this solution into the self-consistency relation at $T=0$ we derive
the following quasilinear equation \cite{VP}:
\begin{equation}
	\pi \Delta(\phi) = \log{(\tau \bar{\epsilon})} \, \hat{V}\Delta(\phi)
		- \log{(\tau  \langle \Delta \rangle)}
			\,\bar{V}(\phi)\langle \Delta \rangle
\label{quasilinear-T=0}
\end{equation}
Assuming that $\log{\tau \bar{\epsilon}} \gg V_0^{-1}$, where $V_0$ is
the the maximal eigenvalue of the operator $\hat{V}$, as previously,
its solution can be found explicitly:
\begin{equation}
	\Delta(\phi) = \langle \Delta \rangle \bar{V}(\phi)/
				\langle \bar{V} \rangle;
	\hspace{0.25in}
	\langle \Delta \rangle = 2 \tau^{\kappa-1}
		\bar{\epsilon}^\kappa \exp(-\pi/\langle \bar{V} \rangle)
\end{equation}
{}From eqns. (\ref{asympt-eta},\ref{asympt-sigma}) it follows that
$\tilde{\eta}(0)=0$ and $\sigma(0)=1/\tau \gg \Delta$.
Since the renormalized OP $\tilde{\Delta}(\phi)$ does not
have any nodes, no divergence can arise in eqn. (\ref{eq:AG-eta}). In this
way a self-consistency of the obtained solution is guaranteed.
\par
Summarizing, $\tilde{\eta}(0) \neq 0$ in a clean superconductor and
it vanishes when a scattering rate substantially exceeds $T_{c0}$.
Hence, there should exist some special value of $\tau = \tau_\star$ at
which $\tilde{\eta}(0)$ first turns into zero. Since $  \left.
\sigma(0) \right|_{\tau = \tau_\star} = 1/\tau_\star $, the value of
$\tau_\star$ can be found as a functional of $\Delta(\phi)$ by means
of the following equation:
\begin{equation}
	\langle \Delta(\phi)/(\Delta(\phi) + 1/\tau_\star) \rangle = 0
\label{tau-star}
\end{equation}
The above equation should be supplemented by the self-consistency and
the AG equations (\ref{eq:self-consistency},\ref{eq:AG-eta},
\ref{eq:AG-sigma})
with $\tau = \tau_\star$. Using this ansatz one can
find $\tau_\star$, at least in principle, as a functional of the OP
in a clean superconductor.
\par
To get more visible results we consider a limiting case $\langle
\bar{V} \rangle^2 \ll\langle \bar{V}^2 \rangle$. If this strong
inequality is satisfied, the average OP should be small compared
to the amplitude of its variation and one can expect that
the dependence $T_c(\tau)$ is close to that for $\langle \Delta
\rangle = 0$.  It means that $T_c(\tau)$ reaches almost zero value as
$\tau \rightarrow \tau_c$, but then has a power-like tail:
$T_c(\tau) \sim \tau^{\kappa-1}$ with $\kappa \gg 1$.  Since the point
$\tau_\star$ separates the domain of $\tau$ with the $d$-like behavior
from that of the $s$-like one, it is natural to conjecture that
$\tau_\star$ is close to $\tau_c$.
The direct calculation justifies this guess.
We assume that $1/\tau_\star \gg \sqrt{\langle \Delta^2 \rangle} \gg
\langle \Delta \rangle$, then eqn. (\ref{tau-star}) yields
\begin{equation}
	\tau_\star = \langle \Delta \rangle / \langle \Delta^2 \rangle
\label{tau-star-delta}
\end{equation}
Plugging the solution of eqn. (\ref{quasilinear-T=0}) into the
expression for $\langle \Delta^2 \rangle$ of the above relation one gets:
\widetext
\begin{equation}
	\tau_\star \langle \Delta \rangle \,
	(\log{(\tau_\star \langle \Delta \rangle})/\pi)^2
	\left \langle \left[
		\left( 1 - \frac{1}{\pi}\log{(\tau_\star \bar{\epsilon})}\,
			\hat{V}\right)^{-1} \bar{V}(\phi)
	\right]^2 \right \rangle
	=1
\label{tau-star-relation}
\end{equation}
\narrowtext
Since $\tau_\star \langle \Delta \rangle $ is assumed to be small,
$\log{(\tau_\star \bar{\epsilon})}/\pi$ must be close to the reciprocal
maximal eigenvalue $V_0^{-1}$ of the operator $\hat{V}$ in order to
satisfy the above equation.
Therefore, $\tau_\star$ coincides with $\tau_c$ for
$\langle \Delta \rangle^2 / \langle \Delta^2 \rangle \rightarrow 0$.
This completes the proof of the statements 2 and 3.
\par
The properties of the AG equations, which have been just established, give us
a clue for an investigation of the behavior of the angular-dependent
DOS.
The latter may be found via the solutions of the AG equations, analytically
continued from the Matsubara sequence, defined on the upper half of the
imaginary axis of the frequency complex plane, to the real axis according
to the formula:
\begin{equation}
\nu_s(\epsilon, \phi) = \nu_n(\epsilon, \phi)   {\cal R}e \left \lbrace
\frac{\tilde{\epsilon}(\epsilon)}{\sqrt{\tilde{\epsilon}^2(\epsilon)-
					\tilde{\Delta}^2(\phi,\epsilon)}}
	\right \rbrace
\label{eq:dens-super-norm}
\end{equation}
Here $\nu_s(\phi)$ and $\nu_n(\phi)$ denote the DOS in the
superconducting and the normal states respectively, and the functions
$\tilde{\epsilon}(\epsilon)$ and $\tilde{\Delta}(\phi,\epsilon)$ represent
the result of the analytical continuation mentioned above. Namely,
$\tilde{\epsilon}(\dot{\imath}\eta_n) = \dot{\imath} \tilde{\eta}_n,\,\,
	\tilde{\Delta}(\phi,\dot{\imath}\eta_n) = \tilde{\Delta}_n(\phi)$.
At the Fermi-level $\tilde{\epsilon}(0)=\dot{\imath} \tilde{\eta}(0)$ for
$ \tau > \tau_\star$ and $\tilde{\epsilon}(0)=0$ for  $ \tau < \tau_\star$.
Hence, $\nu_s(\phi,0)$ is finite for $\tau > \tau_\star$ and vanishes
identically for $ \tau < \tau_\star$. It also vanishes at $\tau=\infty$.
Therefore, it has a maximum at some $\tau > \tau_\star$.
\par
It is natural to suppose, that a gap in the excitation spectrum
does persist for all $\tau < \tau_\star$. We can confirm this conjecture
at least for $\tau \ll \tau_\star$.
The solution (\ref{asympt-eta},  \ref{asympt-sigma}) of the AG equations
can be  employed in this case with the same substitution
$\eta = -\dot{\imath} \epsilon$
and
$\tilde{\eta} = -\dot{\imath} \tilde{\epsilon}$.
Plugging this solution into eqn. (\ref{eq:dens-super-norm}) we find with the
precision up to $\tau\Delta$:
\begin{equation}
\nu_s(\epsilon, \phi) = \nu_n(\epsilon, \phi)   {\cal R}e \left \lbrace
\frac{\epsilon}{\sqrt{\epsilon^2-\langle\Delta\rangle^2}}\right \rbrace
\label{eq:dens-dirty}
\end{equation}
This is exactly the same formula as for an isotropic superconductor
with the average value $\langle\Delta\rangle$ playing the role of the
isotropic gap. This result agrees with the Anderson isotropisation theorem
\cite{anderson}. Note, however, that neither $\nu_s(\epsilon, \phi)$ nor
the OP become isotropic even in the extra dirty limit.
\par
Evaluation with a higher precision shows, that the standard singularity in
DOS (\ref{eq:dens-dirty}) turns into a finite maximum of the height
$
	\sim (\langle\Delta\rangle\tau)^{-1/2}
$,
smeared over the interval $\delta\epsilon \approx \tau\Delta^2$ and
varying with angle. Nevertheless, the threshold character of the DOS
dependence on energy with the threshold approximately
equal to $\langle\Delta\rangle$ remains unaffected.
\par
We have shown that the gap vanishes at $\tau = \tau_\star$ and it also
vanishes at $\tau = 0$. Therefore, the gap reaches its maximum value at some
$\tau < \tau_\star$.
\par
Thus, we propose to look for a rather peculiar phase transition, which can
exist at zero temperature  and is regulated by impurity concentration.
One can try to find it experimentally, scrutinizing the dependence of
quasiparticle tunneling rate or the temperature correction to the
penetration depth on the residual resistivity in single crystals of $YBaCuO$,
doped by $Pr$ or  subjected to the radiation damage. A transmutation of the
gapless tunneling current-voltage characteristics at small resistivity into
the curves with a threshold feature at higher values of residual resistivity
and an analogous conversion of the temperature dependence of the penetration
depth from the gapless to the activated one
would clearly indicate the presence of the transition.
An ideal experimental setting would include the controlling Josephson
tunneling measurements in the corner geometry on the same samples to verify
whether the OP changes its sign.
\par
In conclusion, we have shown that a phase transition should take place at
zero temperature and a special value of the impurity scattering rate
$1/\tau_\star$ in a layered superconductor, provided the OP in the clean
limit does change its sign on the Fermi-surface but its angular average is
finite.
We have argued that, according to the recent
experimental observations, the OP with the required properties is likely
to be an inherent feature, at least for $YBaCuO$.
This transition is characterized by a gap generation in the excitation
spectrum for $\tau < \tau_\star$. The energy gap grows from zero at
$\tau = \tau_\star$ to its maximum at some $\tau > \tau_\star$ and
vanishes again in the extra dirty limit.
The DOS at the Fermi-level turns into zero in the
clean case and at $\tau =\tau_\star$, reaching its maximum at an
intermediate value of $\tau$.
\par
One of the authors (V.L.P) thanks ETH, Zurich, J.~Blatter and
especially D.~Pescia for the hospitality and support extended to him during
the time when this work was in progress.
Thanks are due to H.~Monien and D.~Geshkenbein for discussion and indication
of several references.


\begin{thebibliography}{8}
\bibitem[*]{byline} Present address.
\bibitem{radtke} R.J.~Radtke {\it et al.}, Phys.Rev. {\bf B 48}, 653 (1993).
\bibitem{pines} P.~Monthoux and D.~Pines, Phys.Rev. {\bf B 49}, 4261 (1994).
\bibitem{borkovski}
	L.S.~Borkowski and P.J~Hirschfeld, Phys.Rev. {\bf B49}, 15404 (1994)
\bibitem{Josephson}
	D.A.~Wollman, D.J.~Van~Harlingen, W.C.~Lee, D.M.~Ginsberg and
	A.J.~Legget, Phys.Rev.Lett. {\bf 71}, 2134 (1993);
	D.A.~Brawner and  H.R.~Ott, Phys.Rev. {\bf B50}, 6530 (1994);
	A.~Mathai, Y.~Gim, R.C.~Black, A.~Amar, and F.C.~Wellstood,
	Preprint UMD 1994.
\bibitem{residual}
	J.M.~Valles, A.E.~White, K.T.~Short, R.C.~Dynes, J.P.~Garno,
	A.F.J.~Levi, M.~Anzlovar, and K.~Baldwin, Phys.Rev. {\bf B39}, 11599
	(1989);
	M.B.~Maple {\it et al.}, Journ. Alloys and Compounds {\bf 181}, 135
	(1992);
	A.G.~Sun, L.M.~Paulius, D.A.~Gajewsky, and R.C.~Dynes, Phys.Rev.
	{\bf B50}, 3266 (1994).
\bibitem{sun} A.G.~Sun, D.A.~Gajewsky, M.B.~Maple, and C.R.~Dynes,
	Phys.Rev.Lett. {\bf 72}, 2267 (1994).
\bibitem{Kuboki} K.~Kuboki, M.~Sigrist, Preprint MIT-CMT-9501030;
		K.~Kuboki, P.A.~Lee, Preprint MIT-CMT-9501030,
\bibitem{hohenberg} P.C. Hohenberg, Zh. Exp. Teor. Fiz. {\bf 45},
	1208 (1963),
	[JETP {\bf 18},834 (1964)]
\bibitem{remark}
The root $\tilde{\eta}_0$ has been first discovered by Gor'kov and
Kalugin. See:
L.P.~Gorkov, P.A.~Kalugin,
	Pis'ma Zh. Exp. Teor. Fiz. {\bf 41}, 208 (1985),
	[JETP Lett. {\bf 41}, 263 (1985)].
It was indicated recently
that the AG equations are insufficient in the immediate vicinity of
$\eta = 0$ and non-ladder diagrams should be taken into account,
resulting in nullification of DOS on the Fermi surface.
See:
A.A.~Nersesyan, A.M.~Tsvelik, and F.~Wenger,
	Phys.Rev.Lett {\bf 72}, 2628 (1994).
However, the DOS grows rapidly with energy in the cited model,
reaching Gor'kov-Kalugin's value at some very small scale.
It is not clear whether this psuedo-gap may be detected
experimentally.
\bibitem{VP}
	V.L.~Pokrovsky, Zh. Exp. Theor. Fiz. {\bf 40}, 641 (1961),
	[JETP  {\bf 13}, 447 (1961)].
\bibitem{anderson}
	P.W.~Anderson, J.Phys.Chem.Solids {\bf 11}, 26 (1959).
\end{thebibliography}
\end{document}